\documentclass[11pt,twoside]{article}

\usepackage{asp2006}
\usepackage{epsf}
\usepackage{psfig}
\usepackage{lscape}

\begin{document}
\title{Planetary nebulae as probes of the chemical impact of AGB stars}   
\author{Letizia Stanghellini}   
\affil{National Optical Astronomy Observatory}    

\begin{abstract} 
Planetary nebulae (PN) represent the evolutionary fate of the asymptotic giant branch (AGB)
stellar envelopes, thus 
are ideally suited to study the chemical impact of AGB stars. Stellar evolution predict 
elemental enrichment through the AGB evolution, and convective dredge-up episodes 
allow the products of stellar evolution to reach the stellar outer layers. Planetary nebulae
are probes of these processes, and are also probes of the environment at the time of formation 
of their progenitors, through the elements
not affected by AGB evolution. Ultimately PN may be used to test AGB stars as actors and probes.
Planetary nebulae are easily identified and detected in the galaxy, the Magellanic Clouds, and beyond, thus they 
are probes of AGB evolution an stellar populations in different environments as well.
\end{abstract}

\section{Introduction}  
Planetary nebulae (PN) represent the fate of ejected envelopes of asymptotic giant branch (AGB)
stars, while their central stars (CS) are the remnants of the AGB cores. 
Single (or detached binary) stars with 1$<{\rm M}_{\rm MS}<$8 M$_{\odot}$\footnote{M$_{\rm MS}$ is the mass on
the main sequence} evolve through the 
AGB, and eventually eject their envelope. The central star temperature increases after the 
envelope ejection, until it is high enough for hydrogen to be ionized, giving birth to a PN. 
Planetary nebulae
are characterized by their emission spectra, with the brightest lines corresponding to the Balmer series and
the forbidden [O III] lines, thus they are easy to detect and to identify even in 
outer galaxies. The duration of the PN phase is extremely short if compared with any stellar evolutionary 
phase, since PN evolve on a dynamical rather than nuclear time scale. 
These circumstances contribute to make PN excellent
probes of the recent past of the AGB stars they evolved from.

The spectra of PN may be analyzed relatively easily to determine their chemical abundances. The impact
of PN chemistry is twofold: First, nebular abundances of elements that are expected to
vary with the evolution of low- and intermediate-mass stars allow to use PN as probes of AGB stars 
as actors, and to determine the impact of AGB evolution within the hosting galaxy; second, the study of elements that
are invariant with respect to the evolution of low- and intermediate-mass stars enable PN to be used
as probes of the environment at the time of progenitor formation.

With the current technology, applications concerning the use of PN to determine the chemical impact of AGB stars are 
feasible in the galaxy and the Magellanic Clouds. Beyond that, a few studies of extragalactic PN 
spectroscopy are setting the stage for future endeavors. 
In this paper I review the most recent and homogeneous data sets of PN abundances, including Galactic disk, 
bulge, and Magellanic Cloud PN. A study of these abundances and their comparison to the yields of AGB evolution is
presented in $\S$2, both for Galactic and Magellanic Cloud PN. In $\S$3 I review the use of PN
as probes of stellar populations, including a discussion of the metallicity gradients in the Galactic
disk. In $\S$4 I present the status of PN as AGB probes in other galaxies. Finally $\S$5 includes the 
future challenges of the use of PN as probes of the chemical impact of AGB stars. 

\section{Planetary nebulae as probes of asymptotic giant branch evolution}
Why galaxies care about PN? Planetary nebulae eventually dissolve into the ISM, changing its chemical
composition depending on their own composition. Interestingly, PN may be the main producers of
nitrogen in the galaxy, and compete with massive stars for the production of carbon. Cosmic
recycling is thus the main reason why galaxy care about PN abundances. Furthermore, PN abundances
are a measurable effect of stellar evolution: by appropriately comparing evolutionary yields and 
observed PN abundances one sets the best constraints to stellar evolutionary models. And why should galaxies care 
about stellar evolution models? It turns out that AGB stars are essential contributors to the light in galaxies
(e.g. Maraston 2005), thus understanding which AGB model should be used in population synthesis, 
especially as the metallicity varies, is of great importance for all galaxy studies at all
distances, including those with unresolved stellar populations.

\begin{table}[!ht]
\caption{Galactic disk PN: Average oxygen abundances}
\smallskip
\begin{center}
{\small
\begin{tabular}{lcccc}
\tableline
\noalign{\smallskip}
\noalign{\smallskip}
&& S06 sample& P04 sample& H04 sample\\
\noalign{\smallskip}
\noalign{\smallskip}

\noalign{\smallskip}
\tableline
\noalign{\smallskip}
S06 abundances & O/H [10$^4$]& 3.5 (2.0)& 3.9 (2.2)& 3.7 (1.3)\\
Other abundances & O/H [10$^4$]& & 4.3 (1.9)& 5.6 (1.7)\\

\noalign{\smallskip}
\tableline
\end{tabular}
}
\end{center}
\end{table}

\subsection{Galactic planetary nebulae}  
Abundances in Galactic PN have been obtained by many authors for decades. In the following I 
focus on
recent abundance analyses that include a relatively large number of PN (N$_{\rm PN} > $75), both in the Galactic disk and in the 
bulge. For the Galactic disk, Perinotto et al.~(2004, P04), Henry et al.~(2004, H04), and 
Stanghellini et al.~(2006, S06) have analyzed PN abundances either from original spectroscopy of by reanalysis
of existing data sets. The typical abundance uncertainty in these papers is around 10-15 $\%$ for oxygen, and 
the differences of abundances for the PN in common may be ascribed to the ionization correction factors (ICF) used
in the papers. The averages of the abundance of a given element also differ depending on the
sample. Table 1 gives the average oxygen abundances of the PN in common in the samples of S06 and, respectively, P04 and H04. The first line uses the abundances from S06, while in the second line the
abundances are chosen from the original papers; in this way the columns of Table 1 give the comparison of same samples
of PN with abundances from different studies. The ranges of the averages are given in parenthesis. The difference
between the S06 and H04 samples is, on average, the same than the difference of the ICF used for oxygen abundances.

A recent abundance analysis of Galactic bulge PN is provided by
Exter et al.~(2004, E04), including more than 80 PN. The average abundances of the predominant elements 
in bulge and disk PN are given in Table 2, columns (3) and (4) respectively, and in parenthesis the range of selected diagnostics.
The S06 sample is chosen to represent Galactic disk PN since it explicitly excludes bulge and halo PN, and since the treatment of the ICF therein is the same than
in E04, thus the two samples are directly comparable. Type I PN are those nebulae whose N/O has been enriched with respect to the Orion nebula (E04). 
Carbon abundances for Galactic PN are from Henry et al.~(2000). 
Determination of carbon abundances for bulge PN are sparse, thus their average is not reported.
The data show that the oxygen abundance is higher in bulge than in disk PNe, with an 
offset of about 0.1 dex, while the difference in helium abundance is much lower. 
Nitrogen is also more abundant on average in the bulge, making the N/O ratio almost 
identical in the two environments. 

Figure 1 (left panel) shows the N/O versus O/H locus for the Galactic disk PN analyzed by S06. 
This plot is a good diagnostics to check weather the nitrogen in AGB stars is produced 
by ON cycle. In this case, oxygen would be destroyed to produce nitrogen, and a
strong anti-correlation should be detected in this plot. The Figure
shows a large spread of oxygen abundances, with O/H values much lower than those
used in the overplotted models by Marigo (2001). Oxygen 
depletion occurs as nitrogen is produced, especially in type I PN.
Figure 1 (right panel) shows the diagnostic evolutionary plot of N/O versus He/H again for Galactic
disk PN, clearly indicating that the type I PN correspond to the locus of the 
high mass models (M$_{\rm MS} > 4.5 $ M$_{\odot}$ in Marigo's models).
The C/O versus N/O diagram for Galactic disk PN would be sparsely populated.
Carbon abundances derive from UV lines of IUE spectra, and recent 
carbon determination for disk PN are yet unavailable. 

\begin{figure}[ht!]
\plottwo{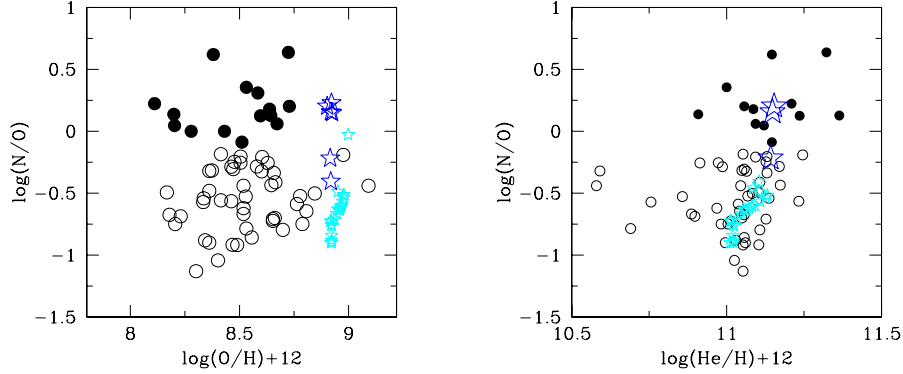}{stanghellini_fig1b.epsi}
\caption{Left panel: log (N/O) versus log (O/H) for Galactic disk PN. 
Open symbols: non-type I PN; filled symbols: type I PN; right panel: log N/O versus He/H. 
Both panels: Small stars indicate models with galactic metallicity from Marigo (2001) with
M$_{\rm MS}< 4.5$M$_{\odot}$; large stars with 
M$_{\rm MS}> 4.5$M$_{\odot}$.}
\end{figure}

\begin{table}[ht!]
\caption{Average abundances in different PN populations}
\smallskip
\begin{center}
{\small
\begin{tabular}{llllll}
\hline
\noalign{\smallskip}
& & Bulge& Disk& LMC& SMC\\
\noalign{\smallskip}
\hline
\noalign{\smallskip}
He/H& 	whole sample& 	0.11 (0.03)& 0.12 (0.04)& 0.10 (0.03)& 0.09 (0.02)\\
    & 	  type I&       0.12& 0.15& 0.11& 0.09\\
    &    non-type I&    0.11& 0.11& 0.10& 0.09\\
    
    &&&&&\\
C/H [10$^4$]&   whole sample &    $\dots$  & 5.7 (6.5)& 3.3 (3.5)& 4.3 (2.5)\\    
&&&&&\\
N/H [10$^4$]&  whole sample   &     2.7 (2.3)& 2.4 (3.5)& 0.94 (0.94)& 0.41 (0.31)\\
&&&&&\\
O/H [10$^4$]&   whole sample  &     4.6 (1.2)& 3.5 (2.0)& 2.1 (1.1)& 0.99 (0.84)\\
& type I&                4.3& 3.1& 1.8& 0.48\\
& non-type I&             4.7& 3.7& 2.4& 1.2\\
&&&&&\\
N/O& 	whole sample&	0.68& 0.66& 0.66& 1.1\\
&type I&   1.1& 1.9& 1.3& 2.9\\
&non-type I& 0.35& 0.32& 0.22& 0.13\\ 
\noalign{\smallskip}
\hline
\end{tabular}
}
\end{center}
\end{table}

\subsection{Magellanic Cloud planetary nebulae}
There are several hundred PN in the Magellanic Clouds (Jacoby 2006), and a good fraction of this 
large population has been studied spectroscopically by several authors 
(Monk et al.~1988; Henry et al.~1989; Boroson \& Liebert 1989; Stasinska et al.~1998). Magellanic Cloud PN abundances are shown, in the form of averages, in Table 2, and 
individually in Figure 2. The carbon abundances of the Magellanic Cloud PN have been derived from UV 
spectra acquired with IUE and HST (Leisy \& Dennefeld 1996; Stanghellini et al.~2005). Type I PN
in the Magellanic Clouds are those whose N/O is enriched with respect to that of the local H II regions.

From Table 2 one infers that the oxygen abundance is, on average, decreasing 
from the galaxy to the LMC and the SMC, as expected. Average oxygen seems to be
depleted in type I PN, indicating that the ON cycle has been active in the PN progenitor. 
Helium abundances is low in LMC and SMC PN. 
Finally, carbon production is less efficient, on average, in the Magellanic Cloud
PN than their Galactic counterparts. The observed carbon depletion might be caused by a more efficient
nitrogen production via the CN cycle. Note that the average carbon abundance of SMC PN in Table 2 is
higher than that of the LMC, but the SMC sample is too small for this average to be statistically meaningful. 
New UV spectra of compact SMC PN 
have been acquired with the prisms of the ACS and will enlarge considerably the sample size of carbon
determinations in SMC PN (Stanghellini et al., in preparation).

\begin{figure}[ht!]
\plotone{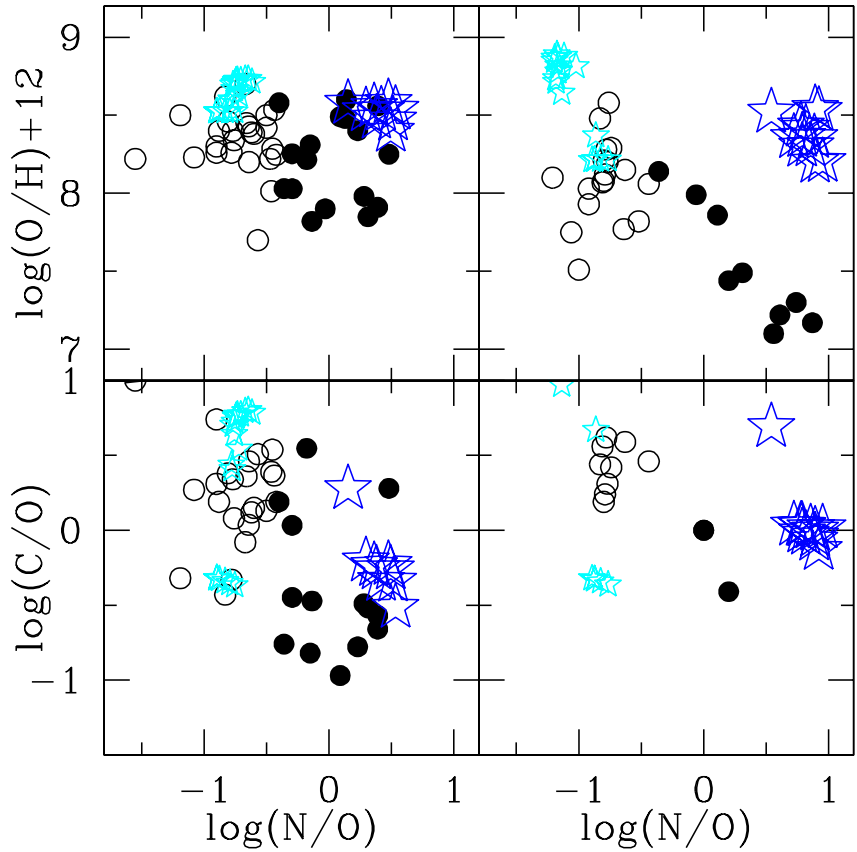}
\caption{Diagnostic diagrams for LMC (left panels) and SMC (right panels) PN. Symbols are as in Fig. 1, right panel.}
\end{figure}

The two upper panels of Figure 2 show the O/H versus N/O plot for Magellanic Cloud PN. 
The depletion of oxygen and the nitrogen enrichment of type I SMC PN is evident (right panel), the same 
effect is milder in the LMC PN. From the comparison of data and models it is inferred
that the ON cycle is much more efficient than predicted by Marigo's (2001) models for the SMC
metallicity. The lower panels of Figure 2 show C/O versus N/O
in LMC (left) and SMC (right) PN. It is worth noting that
in both galaxies the effect of carbon depletion in favor of nitrogen enrichment is
noticeable for type I PN. The LMC and the SMC
AGB models by Marigo (2001) shown in the figure are not adequate to reproduce type I PN,
and show that the nitrogen production is more active at low metallicity than predicted.
A larger sample of carbon abundances should define better the constraints to 
evolutionary models for the SMC PN, and disclose the metallicity dependence of the onset of hot bottom
burning (HBB) and the ON cycle.

\begin{figure}[ht!]
\plotone{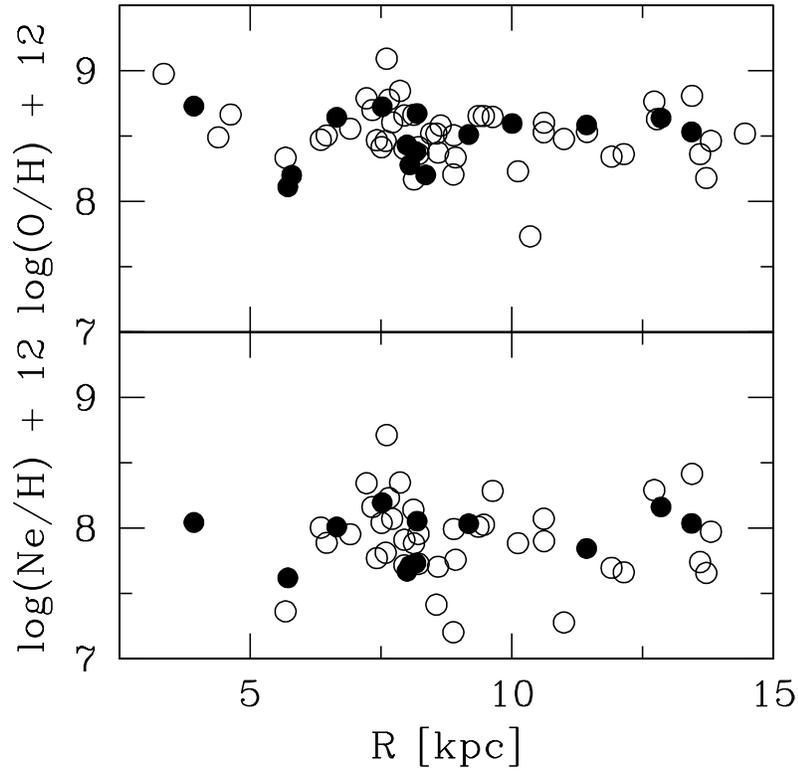}
\caption{Galactic disk PN metallicity gradients, for oxygen and neon. Filled symbols: Type I PN,
open symbols: Non-type I PN.}
\end{figure}

\section{Probing stellar populations}
Planetary nebulae, with their bright spectra and their ubiquity, are also excellent probes
of elements that are not processed during the evolution of their progenitors.
Recent predictions show that neon and sulfur should 
not change during the evolution of PN progenitors, thus their concentration probes the galaxy at the
time of progenitor formation. On the other hand, it is better to exclude type I PN 
when probing Galactic evolution through oxygen abundances, since the HBB or ON cycle may compromise 
the invariance of this element through the late AGB evolutionary stages. 
Stanghellini et al.~(2006) have found that the Galactic disk gradients of PN, evaluated through oxygen and 
neon abundances, are rather flat (of the order of -0.01 dex kpc$^{-1}$ for both elements). 
Figure 3 reports the position of the 
S06 PN on the log R - log (X/H) plane, where type I and non-type I PN are plotted with different symbols. The flatness
of the gradients appear evident. Interestingly, the most recent Galactic disk PN gradient evaluated independently by 
Perinotto \& Morbidelli (2006) agrees with S06. These two recent results are in contrast with 
earlier gradients determination, that would predict higher slopes (up to  -0.07 dex kpc$^{-1}$, Maciel \& Quireza 1999,
Henry et al.~2004). The choices of the ICF are almost irrelevant for the gradient determination, thus
the differences in these values could be ascribed either to the distance scale used for the Galactic PN, or to the
sample of PN analyzed. While several authors conclude that their gradients are independent on the distance scale
(Maciel \& Quireza 1999, Perinotto \& Morbidelli 2006), S06 found that the gradients would steepen if the 
distance scale by Maciel (1984) were used instead of that by Cahn et al.~(1992). On the other hand, since bulge PN
have higher metallicity, it is also important to carefully selected PN that are not in the bulge when
estimating gradients. Last, but not least, one should be aware that all statistical distance scales fail 
for bipolar PN, since these scales are based on the apparent radius measurement. A selection based
on morphological types has been performed by S06, where the gradients are provided for non-bipolar
PN, to avoid statistical distance scale failure.

Figure 4 shows the correlation between neon and oxygen from the PN sample of S06. The two quantities are 
highly correlated, and, as shown in S06, independent on morphological type of the PN 
This is what expected
if both oxygen and neon would derive totally from primary nucleosynthesis, from stars with M$_{\rm MS}>$8 M$_{\odot}$~(a least
square fit to the data would give a slope of 0.34). A careful inspection of Figure 4 shows that the relation splits at low Ne values, and gives lower O/H for type I
than for non-type I PN at the same N/H level. This is once again the signature of a possible ON cycle involving
type I PN, but that does not affect the non-type I PN. 

\begin{figure}[ht!]
\plotone{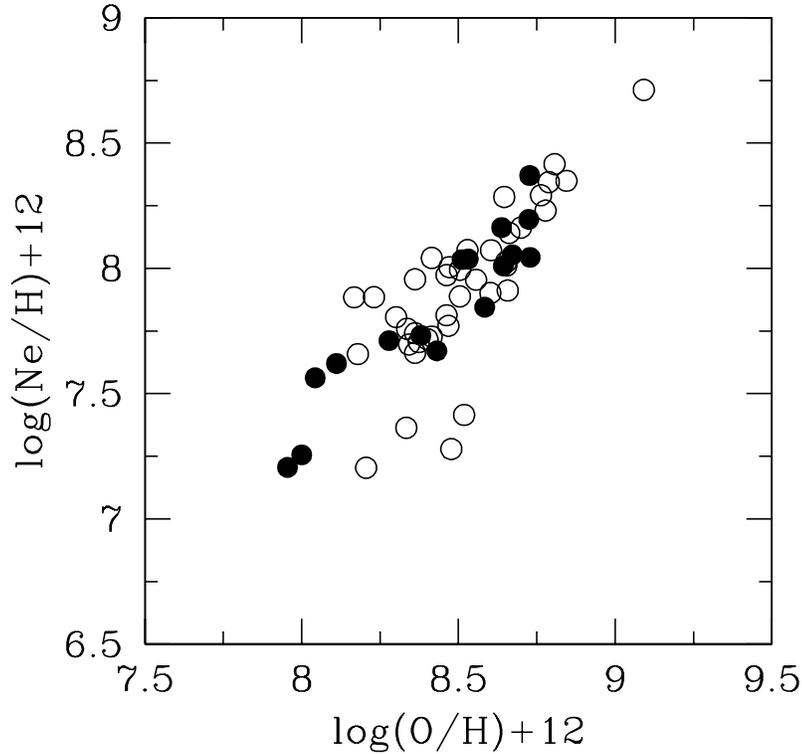}
\caption{The lockstep evolution of neon and oxygen in Galactic disk PN. Filled symbols represent type I PN.}
\end{figure}

\section{Beyond the Magellanic Clouds}
The type of abundance analysis that is reviewed above for the Galactic and Magellanic Cloud PN has been recently extended
to samples of PN at larger distances. A recent review by Richer (2006) summarizes extragalactic PN 
spectroscopic analysis. The detailed diagnostic plots are still limited, but the trends may be compared
with the results of nearby PN. For example, diagnostics of the occurrence of the ON cycle have been 
produced for dwarf spheroidal and dwarf irregular galaxies (Richer et al.~1999; and Richer, private communication).
The neon-oxygen lockstep relation in dwarf galaxies may be reproduced with a collection of spectroscopic data
(Richer et al.~1999; Richer, private communication;
Magrini et al.~2005, Kniazev et al.~2005, van Zee et al.~2006).  

Analysis of PN in spiral galaxies are limited to M31 and M33. Jacoby \& Ciardullo (1999) have explored the 
brightest PN in the M31 bulge and disk, and found that the N/O versus He/H plane is similarly
populated than in Galactic PN. The correlation between neon and oxygen is also available for M31. Magrini et al.~(2004) 
found a hint of a gradient in the oxygen abundance for disk PN in M33, with O/H decreasing inside out the 
galaxy disk, like in the Milky Way. 
Attempts to measure metallicities in elliptical galaxies have so far obtained limited success.
Nonetheless, they show that PN spectra in these galaxies are not dissimilar from those in the Milky Way
(Walsh et al.~1999), and they have disclosed the existence of a metal-rich population in elliptical galaxies 
from the spectra of individual PN (Mendez et al.~2005).

\section{Summary and future challenges}
The chemical impact of AGB stars is very important, both when considering AGB stars as actors, thus
probing stellar evolution, and when using AGB stars as probes of the chemical environment 
at the epoch of progenitor formation. Planetary nebulae trace well the yields of AGB stars, and by comparing yields from
AGB models with PN observations one can constrain the models and disclose the dependence on metallicity of the 
evolutionary processes. Both Galactic and Magellanic Cloud PN have been extensively used for such studies.

In this paper I reviewed chemical abundances analyzed for large PN samples both in the galaxy and the Magellanic 
Clouds. Insight is gained by analyzing these samples and comparing them, taking into account the selection effects
and analysis methodology adopted by individual papers. The evolutionary yields from single star evolution are in broad
agreement with the PN data, with the possible exception of the very low metallicity case, such as the SMC,
should be studied further to have a better model constraint. The elements with zero yield from 
AGB evolution are used to probe stellar populations and the chemical evolution of the galaxy. The latest
Galactic PN studies show that the metallicity gradient is almost flat across the Galactic disk, 
in contrast with previous findings.
This result has broad implications for galaxy evolution studies.

Planetary nebulae beyond the Magellanic Clouds have been recently studied as well, 
these results
are limited to the brightest PN of each galaxy. A future challenge is to explore the realm of 
moderate resolution PN spectroscopy beyond the Magellanic Clouds. 

The models of stellar evolution and their relative yields are built for single stars, while a
fraction of binary stars should be included when comparing data and models. In particular, common envelope
evolution should terminate the AGB phase earlier than for single stars, inevitably changing the yields 
(Izzard et al.~2004). The consequences of including binary AGB evolution is certainly 
a frontier in this field, to be fully explored in the near future.

\acknowledgements 
Thanks to the SOC for the invitation to review this subject, to my collaborators
for their contributions to this talk, to Michael Richer for sharing 
his knowledge on extragalactic PN spectroscopy, and to Corinne Charbonnel, Katia Cunha,
Robert Izzard, Amanda Karakas, John Lattanzio, Eva Villaver, and many others for enjoyable scientific 
discussion during the meeting.

\end{document}